\documentclass[table,conference]{IEEETran}
\IEEEoverridecommandlockouts
\usepackage{algorithmic}
\usepackage{amsmath,amssymb,amsfonts}
\usepackage{booktabs}
\usepackage{caption}
\usepackage{cite}
\usepackage{etoolbox}
\usepackage{float}
\usepackage{graphicx}
\usepackage{highlight}
\usepackage{hyperref}
\usepackage{multirow}
\usepackage{soul}
\usepackage{textcomp}
\usepackage{tcolorbox}
\usepackage{url}

\graphicspath{ {graphics/} }

\renewcommand{\opacity}{50}
\hypersetup{
    colorlinks = true,
	citecolor = {blue},
	urlcolor = {blue},
}

\def\BibTeX{{\rm B\kern-.05em{\sc i\kern-.025em b}\kern-.08em
    T\kern-.1667em\lower.7ex\hbox{E}\kern-.125emX}}

\begin{document}

\title{Syndeo: Portable Ray Clusters with Secure Containerization}

\author{
	\IEEEauthorblockN{William Li}
	\IEEEauthorblockA{
		\textit{MIT Lincoln Laboratory} \\
		Lexington, MA, USA \\
		william.li@ll.mit.edu
	}
	\and
	\IEEEauthorblockN{Rodney S. Lafuente Mercado}
	\IEEEauthorblockA{
		\textit{MIT Lincoln Laboratory} \\
		Lexington, MA, USA \\
		rodney.lafuentemercado@ll.mit.edu
	}
	\and
	\IEEEauthorblockN{Jaime D. Pe\~na}
	\IEEEauthorblockA{
		\textit{MIT Lincoln Laboratory} \\
		Lexington, MA, USA \\
		jdpena@ll.mit.edu \\
	}
	\and
	\IEEEauthorblockN{Ross E. Allen}
	\IEEEauthorblockA{
		\textit{MIT Lincoln Laboratory} \\
		Lexington, MA, USA \\
		ross.allen@ll.mit.edu
	}
}

\maketitle

\begin{abstract}
	We present Syndeo: a software framework for container orchestration of Ray on Slurm.  In general the idea behind Syndeo is to \textit{write code once and deploy anywhere}.  Specifically, Syndeo is designed to addresses the issues of \textit{portability}, \textit{scalability}, and \textit{security} for parallel computing.  The design is portable because the containerized Ray code can be re-deployed on Amazon Web Services, Microsoft Azure, Google Cloud, or Alibaba Cloud.  The process is scalable because we optimize for multi-node, high-throughput computing.  The process is secure because users are forced to operate with unprivileged profiles meaning administrators control the access permissions.  We demonstrate Syndeo's portable, scalable, and secure design by deploying containerized parallel workflows on Slurm for which Ray does not officially support.\footnote{DISTRIBUTION STATEMENT A. Approved for public release. Distribution is unlimited.  This material is based upon work supported by the Department of the Air Force under Air Force Contract No. FA8702-15-D-0001. Any opinions, findings, conclusions or recommendations expressed in this material are those of the author(s) and do not necessarily reflect the views of the Department of the Air Force.}
\end{abstract}


\section{Introduction}

The availability of High Performance Computing (HPC) has revolutionized Artificial Intelligence (AI) research in a wide variety of domains\cite{gozalobrizuela2023survey}\cite{DBLP:journals/corr/abs-1708-05866}. However, the computing costs necessary for these types of algorithms are quite extraordinary. For example, DeepMind's computing cost needed to train their Go algorithm reached upwards of \$25M\cite{alphago-cost}. This has created a tremendous need for developing hardware infrastructure that enables efficient computation, and, similarly, schedulers that enables massively parallelized workflows.

Organizations generally have two options when building hardware infrastructure:  \textit{cloud-based} or \textit{on-premises} infrastructure. Cloud-based infrastructure provide organizations with the ability to scale up their compute on demand, but may be prohibitively expensive compared with on-premesis solutions\cite{Fisher2018}. In contrast, on-premises solutions allow organizations full control of their computing resources but puts the onus of maintenance on the organization itself.

In addition to hardware infrastructure, organizations also need \textit{schedulers} for distributing jobs across their computing infrastructure.  \textit{Scheduling} is the process of assigning hardware/software resources to perform specific jobs or tasks.  It is important to keep in mind that schedulers are chosen based on the specific needs of an organization or user.  Within the scientific research community, there is often an on-premesis shared HPC serving multiple developers.  These are often referred to as \textit{multi-tenant} environments.  But in the commercial world, developers can immediately request on-demand computing resources from a cloud provider for individual use.  These are referred to as \textit{single-tenant} environments.  Because these communities have different needs, they often choose different schedulers.

But different schedulers utilize different syntax meaning that code written for one scheduler is often incompatible with a different scheduler.  Among the scientific community, Slurm has emerged as the de-facto standard based on the fact that approximately 60\% of the top 500\cite{USC-slurm-desc} most powerful non-distributed computer systems in the world use it.  Conversely, Ray has emerged as popular open-source scheduler widely used for cloud computing in the commercial world.  Note that these two schedulers are incompatible with each other as Slurm was designed for multi-tenant on-premesis HPC scheduling whereas Ray focuses on single-tenant cloud computing.

Furthermore, scheduler incompatibility is not the only issue for portability across different HPC systems.  Software deployment also needs to be \textit{containerized} in order to work on different architectures.  Here we introduce another problem between multi-tenant environments versus single-tenant environments.  In multi-tenant environments users must have \textit{unprivileged} access whereas single-tenant environments often have \textit{privileged} access.  This means that different environments often use different types of containers.  To make the example more concrete, multi-tenant scientific research environments often use Apptainer/Singularity\cite{singularity-paper} for containerization whereas commercial environments often use Docker\cite{10.5555/2600239.2600241}.  The main difference between these two container types is that Apptainer/Singularity provides security assurances whereas Docker does not.  These two types of containerization solutions are incompatible with each other and presents another problem for cross-compatibility.

Porting code from Slurm to Ray or vice versa would normally require a rewrite of both the schedulers and the containers.  This may be feasible for smaller projects but would not work for larger projects spanning millions of lines of code.  We exemplify the benefit of using Syndeo and its cross-domain features by running Ray, a scheduler incompatible with Slurm, on Slurm.  We re-emphasize that, while we showcase its use in Slurm, it is usable with other cloud-based or on-premises hardware infrastructures.

The need for a cross-compatible scheduling solution that is deployable across any computing architecture with security assurances is both a challenge and an opportunity for academic and commercial industries alike.  Syndeo presents a solution that is scalable, secure, and cross-compatible for on-premises or cloud computing \textit{without rewriting the software}.  The central idea behind Syndeo is to write software once and scale on any computing architecture with minimal effort.

We present additional background on Slurm, Ray, and containerization solutions in section \ref{background}. Our methodology for developing Syndeo is described in section \ref{sec:methodology}. Section \ref{sec:experiments} presents experimental results, with concluding remarks in Section \ref{sec:conclusion}.  The full Syndeo repository can be found at: https://github.com/mit-ll/Syndeo.

\section{Background}\label{background}

\subsection{Slurm}
Slurm also known as Simple Linux Utility for Resource Management, is an open-source job scheduler for Linux and Unix-like kernels.  The development of Slurm began at Lawrence Livermore National Laboratory in 2002\cite{Jette2023}.  Originally designed as a simple resource manager, it has grown its features and capabilities over time.  Today Slurm is widely used at government institutions, universities, and companies worldwide \cite{slurm_klusacek}. Over half\cite{USC-slurm-desc} of the top 500 most powerful non-distributed computer systems in the world use Slurm.

Even with the rise of cloud computing, Slurm still offers a compelling proposition for on-premesis HPC.  Research by Fisher\cite{Fisher2018} has shown that cloud infrastructure often costs twice as much as on-premesis infrastructure over 10-year cycles.  With cloud infrastructure you are paying for the luxury of \textit{convenience}.  The cost of convenience is small initially but rises over time, a quote by Fisher is appropriate here:

\begin{tcolorbox}
	If an organization has high confidence in the capacity of internal IT resources and high confidence in their ability to deliver necessary results, then the On-Premise cost structure can be expected to save money over the longer term when compared to Cloud. If, instead, the convenience and flexibility of Cloud is sought, with the extra hand-holding for upgrades and guidance services, and the operating budget can afford an ongoing multi-year Cloud subscription, then relying on the external centralized hosted solution could be more compelling\dots

	Electing to choose Cloud is a luxury that allows the organization to avoid incurring direct internal costs such as infrastructure headcount, hardware operations, system administration, etc.
\end{tcolorbox}

Despite the increased costs, commercial industries have gravitated to solutions like Amazon Web Services (AWS), Microsoft Azure, Google Cloud, or Alibaba Cloud.  Even though the major cloud providers offer support for Slurm integration on their infrastructure, Slurm is rarely used on the cloud.  The reasons are two-fold: (1) Slurm is more cost-efficient when paired with on-premesis compute, and (2) the commercial world has developed new schedulers specifically designed to operate on cloud infrastructure.

\subsection{Ray}
One of the most widely used open-source schedulers that is compatible with the major cloud providers is \textit{Ray}\cite{DBLP:journals/corr/abs-1712-05889}.  Ray is an open-source scheduler designed for Python based workflows which originated from researchers at Berkley University.  Code written in Ray is cross-compatible across all four major cloud providers.  Ray was also designed with many modern software features conducive for AI research\cite{Pumperla_Liaw_Oakes_2023}.  Due to Ray's rich feature set, cloud compatibility, and detailed documentation, the scheduler has been widely adopted by the open-source community.  Unfortunately, Ray does not officially support Slurm.  Their rational is restated here\cite{ray-slurm-support}:

\begin{tcolorbox}
	\begin{itemize}
		\item SLURM requires multiple copies of the same program are submitted multiple times to the same cluster to do cluster programming. This is particularly well-suited for MPI-based workloads.

		\item Ray, on the other hand, expects a head-worker architecture with a single point of entry. That is, you will need to start a Ray head node, multiple Ray worker nodes, and run your Ray script on the head node.
	\end{itemize}
\end{tcolorbox}

Because of this contention between the two schedulers, they are not compatible with each other.  The Ray team has stated that they are not interested in officially supporting Slurm.  Slurm researchers would essentially need to rewrite their code if they wanted to operate on Ray.  This is not feasible for larger software projects.  Furthermore, Ray was designed with AI workflows in mind and has its own vibrant research community.  Without a cross-compatible solution researchers using Slurm would be cut off from many modern AI workflows developed under Ray.

According to GitHub as of March 2024 the Slurm project only has 2.2K\cite{github-slurm} stars whereas Ray has nearly 30K\cite{github-ray} stars.  If these numbers are representative of the open source community's support for these scheduler types, it would mean that Ray's community is approximately 13x larger than Slurm's community.  When comparing open-source software it is important to consider the ecosystem of developers supporting the software.

In this comparison, Ray has a larger community with a larger support base that is contributing new features to the project on a daily basis.  Many of the modern AI workflows are being built on Ray instead of Slurm.  Without the ability to port code from Ray to Slurm and vice versa, Slurm researchers will be at a severe disadvantage compared to researchers leveraging Ray.  This is one of the core issues that Syndeo attempts to solve.

\subsection{Containerization - Docker vs. Apptainer}
But in order to truly make code cross-compatible across Slurm and commercial cloud providers, \textit{containerization} is needed for the underlying software even if the scheduling problem is addressed.  Containerization is the process of isolating programs to run in user spaces called \textit{containers}.  Advantages of containerization include: code isolation, portability, and maintainability.  Containers are different from \textit{virtual machines} which offer complete kernel isolation emulating their own hardware and OS (Operating System).  Containers utilize the same kernel and OS of the host system making them more compute efficient than virtual machines.

In the commercial world, Docker\cite{10.5555/2600239.2600241} is the most widely adopted containerization software of choice due to its vibrant user community, good performance\cite{container-speed-test}, and deep documentation.  Unfortunately Docker has security issues involving root escalation which has prevented it from wide adoption in the scientific computing community.  A detailed description of all of Docker's security vulnerabilities is beyond the scope of this paper, but cybersecurity companies like Snyk have posted\cite{docker-security-issues} issues with Docker. We include a quote from Kurtzer\cite{singularity-paper} that aptly describes the problem:

\begin{tcolorbox}
	While Docker takes steps to mitigate the risk of allowing users to run arbitrary code, there is a fatal design flaw that limits Docker’s ability to run in HPC environments: for every container that Docker runs, the container process is spawned as a child of a root owned Docker daemon. As the user is able to directly interact with and control the Docker daemon, it is theoretically possible to coerce the daemon process into granting the users escalated privileges. Any user being able to escalate up to system administrator status, a user called “root”, would introduce unthinkable security risks for a shared compute environment.
\end{tcolorbox}

Schedulers like Slurm, which operate in a multi-tenant environment, cannot use Docker containers because of the security issues just described.  In general, administrators of multi-tenant environments need the ability to limit the range of permissible actions and data access of their users.

Docker's security issues were well known and researchers began looking at ways to address these issues.  In 2017 Abdulrahman\cite{7923813} presented a method to enforce the execution of Docker containers as the submitting user instead of root.  This opened the way for researchers to successfully deploy Docker containers on to multi-tenant systems like Slurm.

In parallel to Abdulrahman's efforts, in 2017 a group of researchers at Lawrence Berkeley National Laboratory led by Gregory Kurtzer developed \textit{Singularity}\cite{singularity-paper}.  Singularity was a new containerization solution built from the ground up designed with security in mind.  A comparison of Singularity versus Docker's security is summarized on Apptainer's website\cite{singularity-security}:

\begin{tcolorbox}
	A default Docker installation uses a root-owned daemon to start containers. \dots
	since the daemon is privileged, users can ask it to carry out actions that they wouldn't normally have permission to do.

	When a user runs a container with Singularity, it is started as a \textbf{normal process running under the user's account}. Standard file permissions and other security controls based on user accounts, groups, and processes apply.
\end{tcolorbox}

Besides the security differences, there are also differences in how these container architectures were designed.  Docker was primarily designed as a container for hosting \textit{microservices} whereas Singularity was designed for running heavy computational jobs over long periods of time, also known as \textit{high throughput computing}.  Kubernetes had this to say about microservices versus high throughput applications\cite{kubernetes-hpc}:

\begin{tcolorbox}
	HPC applications are often either optimized towards massive throughput or large scale parallelism. In both cases startup and teardown latencies have a discriminating impact. Latencies that appear to be acceptable for containerized microservices today would render such applications unable to scale to the required levels.
\end{tcolorbox}

Singularity was designed specifically to address needs of the HPC community.  Features include: high throughput computing, support for multi-tenancy, support for multiple schedulers, designed for general scientific use, immutability of containers for portability, etc.  These design decisions make Singularity uniquely suited for high-throughput applications in contrast to Docker.  In 2021 the Singularity open source project joined the Linux Foundation and was renamed \textit{Apptainer}\footnote{Apptainer is sometimes referred as Singularity.  This is because Singularity was the original name of the container project and Apptainer became an open-source fork of Singularity.}.  It is because of the reasons stated above that Syndeo utilizes Apptainer/Singularity instead of Docker as its primary containerization solution.

\section{Methodology}\label{sec:methodology}

\subsection{Slurm Scheduling vs. Ray Scheduling}
Before discussing Syndeo, we first need to have an overview of how Slurm performs scheduling and how Ray performs scheduling.  Once we have an understanding of how these different schedulers work we will then introduce the unique configuration of how Syndeo operates.  Here we assume that we have a Python script that has a parallel component which we are about to submit for processing.

\begin{figure}
	\centering
	\includegraphics[width=\linewidth]{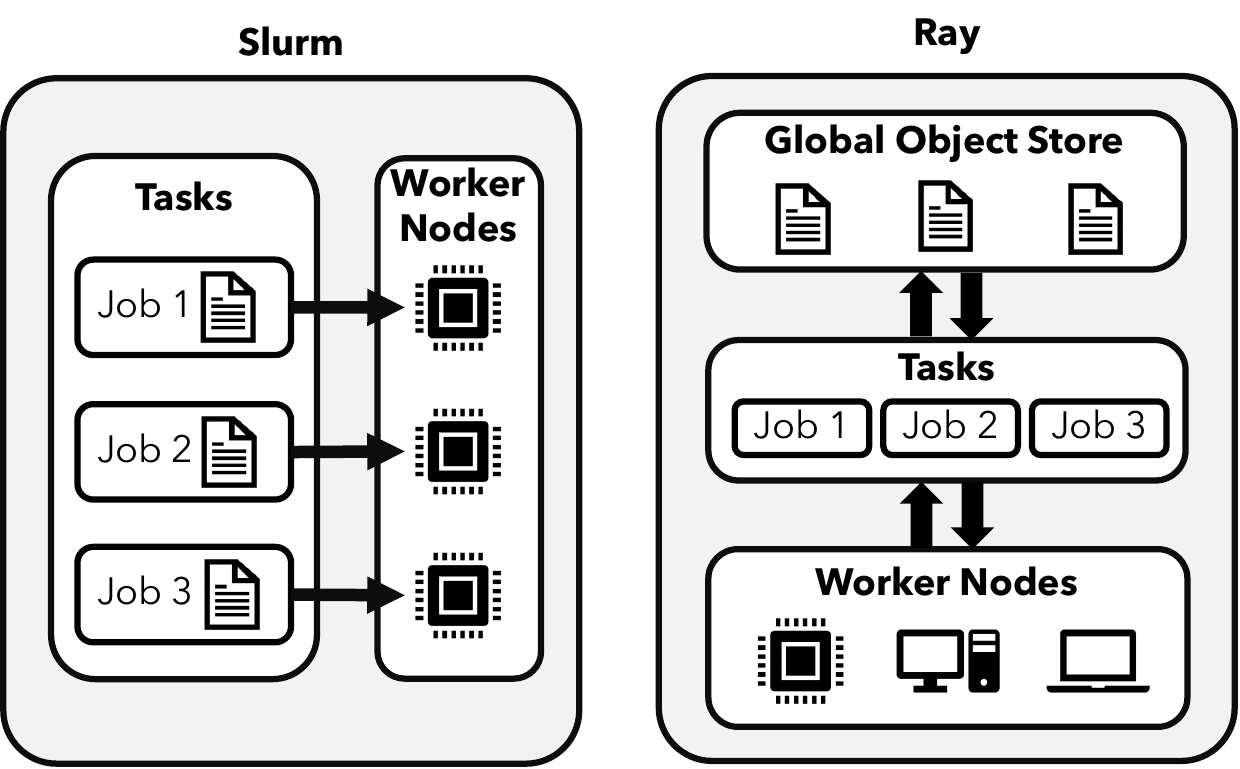}
	\caption{In the Slurm paradigm jobs are sent to homogeneous worker nodes with the data it needs to process.  When the worker nodes are done processing their jobs, synchronization points are used to aggregate the data.  Ray operates with a different paradigm when running jobs.  In Ray, each job is an abstraction that cannot start unless all of its dependencies are met.  Dependencies can be sourced from heterogeneous worker nodes (computing resources) or data from the Global Object Store.  Jobs can push artifacts to the Global Object Store which may be dependencies for other jobs.
		\emph{Note that this is a simplified description of Slurm and Ray.  Slurm offers multiple plugins which can change how it processes jobs.}}
	\label{fig:slurm-vs-ray}
\end{figure}

In Slurm (see Figure~\ref{fig:slurm-vs-ray}, left), the scheduler begins by allocating worker nodes for a set duration.  The centralized manager then dispatches jobs to the worker nodes.  Each worker node has a Slurm daemon, which waits for work, executes that work, returns status, and waits for more work.  In the Slurm paradigm, the common design pattern is to submit multiple copies of the same algorithm to each node and have them process different parts of the input data.  This is conceptually easy to understand and program for because each worker node is given the same set of instructions, only their input data differs.

Ray operates with a different design pattern (see Figure~\ref{fig:slurm-vs-ray}, right).  Ray expects a \textit{head-worker} architecture.  To put it another way, Ray starts a head node, with multiple Ray worker nodes, and expects your script to run on the \textit{head node}.  The configuration of a Ray head node with one or more Ray workers is called a \textit{Ray Cluster}.  In a Ray Cluster configuration, every job is an abstraction that simply needs to meet its \textit{dependencies}\footnote{Dependencies can be either physical resources like Computer Processing Units (CPU)/Graphical Processing Units (GPU); or they can be data dependencies like files, logs, etc.} in order to start.

Ray jobs get their data dependencies from the \textit{Global Object Store}.  Ray jobs may produce \textit{artifacts} which can be sent to the Global Object Store.  Downstream Ray jobs may require certain artifacts as dependencies.  Recall that dependencies can either be physical resources like CPUs or data artifacts.  Ray does not make a distinction as all dependencies must be satisfied before running a job.  This type of dynamic scheduling is generally less efficient than Slurm's paradigm of assigning a job to each worker node.  However, Ray's scheduling paradigm does provide more flexibility for certain workflows where jobs are dynamically allocated.

\subsection{Ray Scheduler inside Slurm Scheduler}
Because Slurm and Ray operate differently, code designed to run on one architecture will be incompatible on the other.  Syndeo solves the Slurm/Ray incompatibility issue by standardizing around Ray's coding syntax and operating only with the Ray scheduler.  At this point the reader may object and state that the Ray scheduler is incompatible with the Slurm scheduler.  The key to making Ray work on Slurm is to set up a \textit{scheduler inside a scheduler}.  This is an unusual concept and needs further explanation.

Syndeo builds a Ray Cluster within Slurm and send its jobs to the Ray Cluster instead of sending them to Slurm.  The Ray Cluster is unaware of the Slurm scheduler; as long as the Ray Cluster can communicate among its nodes over Internet Protocol (IP) it can operate independently.  This is what is meant by setting up a \textit{scheduler inside a scheduler}.

\subsection{Ray Containerization}

Syndeo leverages Apptainer/Singularity to containerize the Ray Cluster, environment variables, libraries, algorithms, and other runtime executables.  Every node that wants to join the Ray Cluster must have a copy of the container.  But having a copy of the container at every node isn't enough, there also needs to be networking among the various nodes in order to bring up the Ray Cluster.  Syndeo performs networking coordination among the various nodes when connecting the containers together when operating with Slurm.

\subsection{Slurm Hosting Ray Cluster}\label{sec:syndeo}

\begin{figure*}[htbp]
	\centering
	\includegraphics[width=\linewidth]{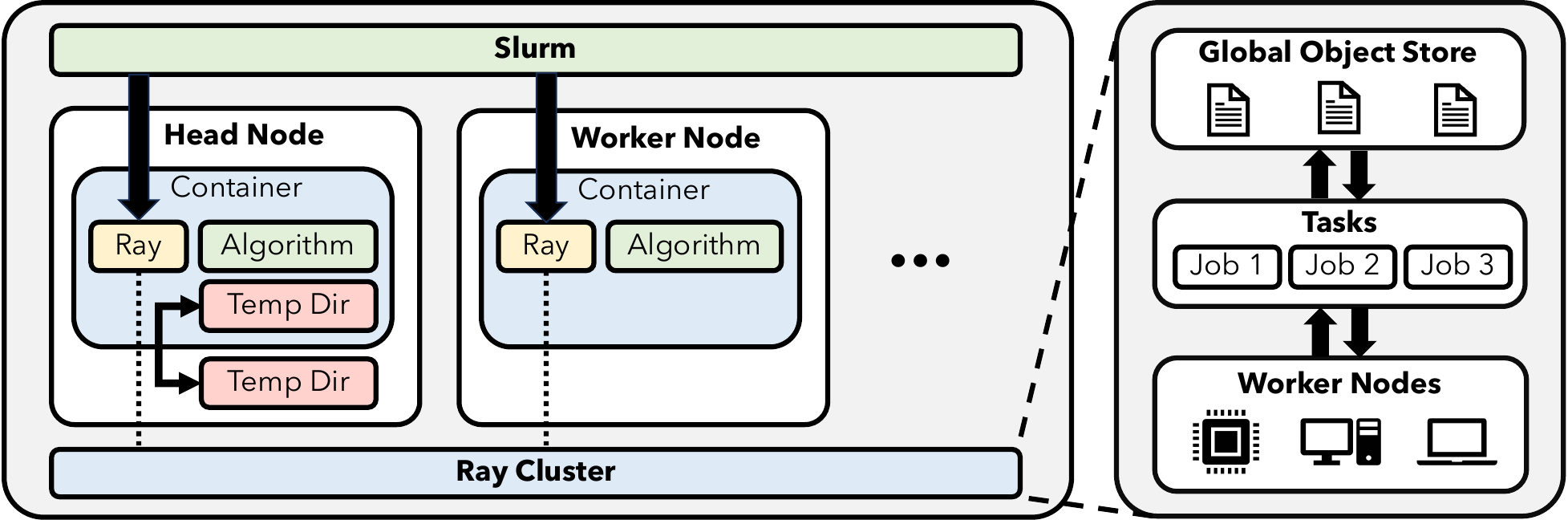}
	\caption{Syndeo starts container orchestration by providing a copy of the container to all allocated nodes.  One head node will be assigned and the rest will be worker nodes.  Each container is preconfigured with Ray and the user's algorithm(s).  At runtime, Syndeo initializes Ray on all containers and checks for network connectivity.  If all the Ray containers successfully connect, they form a \textit{Ray Cluster}.  The Ray Cluster allows users to submit jobs and will execute them on the Ray scheduler.  Syndeo offers a simple container orchestration method that is compatible with Slurm.}
	\label{fig:syndeo-overview}
\end{figure*}


The order of operations needed to bring up a Ray Cluster on Slurm (Figure~\ref{fig:syndeo-overview}) or cloud provider can be broken down into four phases: \textit{creating the container}, \textit{starting the Ray head}, \textit{adding Ray workers}, and \textit{running the Ray Cluster}.  These four phases involve multiple sub-steps that must be executed properly in order for the Ray Cluster to set up correctly.

\begin{enumerate}
	\item \textbf{Creating the Container} - At this stage, the user builds and verifies that Ray can run the user's algorithm inside a self-isolated container on a single node.  This needs to be done on a development machine where the user has root access because the process of building containers needs privileged access.  After the container is built and tested, it needs to be copied over to the Ray head node as well as the Ray worker nodes.
	\item \textbf{Starting the Ray Head} - The Ray head node's IP address and port number are passed into the container to start Ray initialization.  During the initialization process, Ray needs to bind to a temporary directory with read/write access on the host system\footnote{The Ray temporary directory is used to read/write files necessary for its object store functions and scheduling processes.}.  We also turn on Apptainer's \textit{sandboxing} feature which allows it to write internally within its container without binding to the external file system.

	      In order for other Ray nodes to connect to the Ray head container, we need to write its IP address and port number to a shared location where other Ray nodes can access that information.  For Slurm, this would be a shared file system.  For a cloud provider it will be a service (i.e. AWS Simple Storage Service).

	\item \textbf{Adding Ray Workers} - The IP address and port number of the Ray head node is passed into the worker node's container as part of the Ray initialization process.  During container initialization we activate Apptainer's \textit{sandboxing} feature to ensure writing is possible within the container.  Some handshaking protocols occur to verify the communication between the Ray head and Ray workers.
	\item \textbf{Running the Ray Cluster} - Once the Ray head and Ray workers verify their network communications the Ray Cluster is established.  At this point, all the Ray nodes are operating within their containers isolated across different nodes.  The Ray Cluster communicates among its various nodes using network IP and port numbers.  The Ray Cluster is ready to start accepting jobs at the head node using a \textit{head-worker} architecture.
\end{enumerate}

\subsection{Cloud Scheduling}
When deploying to the cloud Apptainer/Singularity offers \textit{container orchestration} through Kubernetes.    Kubernetes is an open source container orchestration tool that was originally developed and designed by engineers at Google.  Today Kubernetes is the de-facto standard with over 90\% of the market share\cite{kubernetes-market} in the commercial space.  When running on the cloud, Kubernetes manages the Apptainer/Singularity pods in a network configuration.  A containerized Ray Cluster can be formed by networking all the individual containers together making your code cross-compatible for public or government cloud architectures.  However, Kubernetes does not operate on Slurm, which is why Syndeo is needed.


\subsection{Repository}\label{sec:repository}

All code for this paper can be found on GitHub: \href{https://github.com/mit-ll/Syndeo}{Syndeo}, \href{https://github.com/mit-ll/RL-Benchmarks}{RL Benchmarks}, and \href{https://github.com/mit-ll/Apptainer-Templates}{Apptainer Templates}.  Full documentation is hosted on GitHub Pages.  Pytests have been provided to verify that the code runs on the Slurm architecture.  The experimental results presented in this paper can be replicated using the Apptainer definition files.  These will build a container that fully replicates all the code, environments, and configuration settings used in this paper.

\section{Experiments}\label{sec:experiments}

\begin{figure*}
	\centering
	\includegraphics[width=\linewidth]{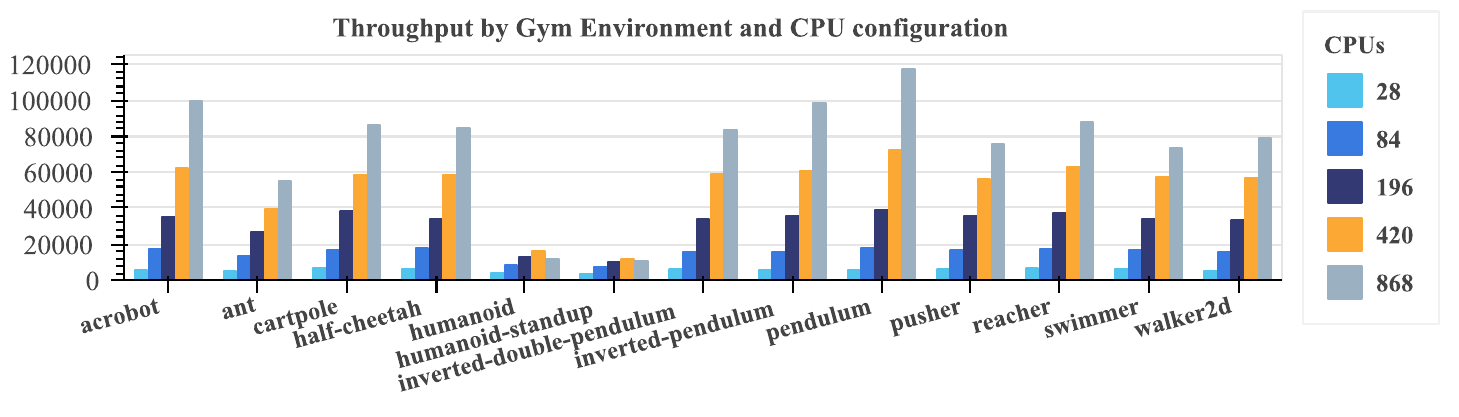}
	\caption{\label{fig:bar}All environments tested with their corresponding CPU configuration and throughput values.}
\end{figure*}

In order to benchmark Syndeo we configured Ray Clusters with varying number of CPU workers with an on-premesis Slurm architecture.  Each CPU worker is an Intel Xeon E5-2683 2.0Ghz.  Each node consisted of 28 CPU workers and 256 GB Random Access Memory (RAM). All nodes are connected by a non-blocking 10-Gigabit Ethernet network and a non-blocking Intel OmniPath low-latency network.  The shared central storage system uses a 10 petabyte Seagate/Cray ClusterStor CS9000 storage array that is directly connected to the core switch.  Each CPU worker was assigned a Ray actor which hosted a Gymnasium\cite{towers_gymnasium_2023} environment and a fully connected neural network.  The environments benchmarked were a combination of classic control simulations\cite{gym-acrobot, gym-cartpole, gym-pendulum} and MuJoCo\cite{todorov2012mujoco, gym-ant, gym-half-cheetah, gym-hopper, gym-humanoid, gym-humanoidstandup, gym-inverted-double-pendulum, gym-inverted-pendulum, gym-pusher, gym-reacher, gym-swimmer, gym-walker2d} simulations (see Figure~\ref{fig:bar}).

\begin{table}
	\centering
	\fontfamily{ppl}\selectfont
	\begin{tabular}{r|lllll}
		\toprule
		\multicolumn{6}{c}{\textbf{Throughput Speedup Factors}}                                                                                                                                                               \\
		\toprule
		                                                                                                           & \multicolumn{5}{c}{CPUs (Workers)}                                                                       \\
		Environment                                                                                                & 28                                 & 84             & 196            & 420             & 868             \\
		\toprule
		\href{https://gymnasium.farama.org/environments/classic_control/acrobot/}{Acrobot}                         & \gradientB{1}x                     & \gradientB{3}x & \gradientB{6}x & \gradientB{11}x & \gradientB{18}x \\
		\href{https://gymnasium.farama.org/environments/mujoco/ant}{Ant}                                           & \gradientB{1}x                     & \gradientB{3}x & \gradientB{5}x & \gradientB{8}x  & \gradientB{11}x \\
		\href{https://gymnasium.farama.org/environments/classic_control/cart_pole/}{Cartpole}                      & \gradientB{1}x                     & \gradientB{2}x & \gradientB{6}x & \gradientB{8}x  & \gradientB{13}x \\
		\href{https://gymnasium.farama.org/environments/mujoco/half_cheetah}{Half Cheetah}                         & \gradientB{1}x                     & \gradientB{3}x & \gradientB{5}x & \gradientB{9}x  & \gradientB{13}x \\
		\href{https://gymnasium.farama.org/environments/mujoco/hopper}{Hopper}                                     & \gradientB{1}x                     & \gradientB{3}x & \gradientB{6}x & \gradientB{10}x & \gradientB{16}x \\
		\href{https://gymnasium.farama.org/environments/mujoco/humanoid}{Humanoid}                                 & \gradientB{1}x                     & \gradientB{2}x & \gradientB{3}x & \gradientB{4}x  & \gradientB{3}x  \\
		\href{https://gymnasium.farama.org/environments/mujoco/humanoid_standup}{Humanoid Standup}                 & \gradientB{1}x                     & \gradientB{2}x & \gradientB{3}x & \gradientB{3}x  & \gradientB{3}x  \\
		\href{https://gymnasium.farama.org/environments/mujoco/inverted_double_pendulum}{Inverted Double Pendulum} & \gradientB{1}x                     & \gradientB{2}x & \gradientB{5}x & \gradientB{9}x  & \gradientB{13}x \\
		\href{https://gymnasium.farama.org/environments/mujoco/inverted_pendulum}{Inverted Pendulum}               & \gradientB{1}x                     & \gradientB{3}x & \gradientB{6}x & \gradientB{10}x & \gradientB{17}x \\
		\href{https://gymnasium.farama.org/environments/classic_control/pendulum/}{Pendulum}                       & \gradientB{1}x                     & \gradientB{3}x & \gradientB{7}x & \gradientB{12}x & \gradientB{20}x \\
		\href{https://gymnasium.farama.org/environments/mujoco/pusher}{Pusher}                                     & \gradientB{1}x                     & \gradientB{3}x & \gradientB{6}x & \gradientB{9}x  & \gradientB{13}x \\
		\href{https://gymnasium.farama.org/environments/mujoco/reacher}{Reacher}                                   & \gradientB{1}x                     & \gradientB{3}x & \gradientB{6}x & \gradientB{10}x & \gradientB{13}x \\
		\href{https://gymnasium.farama.org/environments/mujoco/swimmer}{Swimmer}                                   & \gradientB{1}x                     & \gradientB{3}x & \gradientB{6}x & \gradientB{9}x  & \gradientB{12}x \\
		\href{https://gymnasium.farama.org/environments/mujoco/walker2d}{Walker2d}                                 & \gradientB{1}x                     & \gradientB{3}x & \gradientB{6}x & \gradientB{11}x & \gradientB{15}x \\
		\bottomrule
	\end{tabular}
	\newline
	\caption{\label{tab:experiment-speedup}Throughput speedups based on different CPU worker configurations.}
\end{table}

The Ray actors interacted with the environment by taking state inputs and generating output actions analogous to an AI agent interacting with its environment.  To saturate the CPUs on the Ray Cluster we set the number of state-actions interactions as 1,000 multiplied by the number of CPU cores in the Ray Cluster.\footnote{For example, a Ray Cluster with 28 CPU workers would be tasked with collecting 28,000 samples whereas a Ray Cluster with 868 CPU workers would be tasked with collecting 868,000 samples.}  Five unique CPU configurations were tested (28, 84, 196, 420, and 868).

\begin{table}
	\centering
	\fontfamily{ppl}\selectfont
	\begin{tabular}{r|lllll}
		\toprule
		\multicolumn{6}{c}{\textbf{Throughput Efficiency Percentages (\%)}}                                                                                                                                                  \\
		\toprule
		                                                                                                           & \multicolumn{5}{c}{CPUs (Workers)}                                                                      \\

		Environment                                                                                                & 28                                 & 84              & 196            & 420            & 868            \\
		\toprule
		\href{https://gymnasium.farama.org/environments/classic_control/acrobot/}{Acrobot}                         & \gradientA{100}                    & \gradientA{100} & \gradientA{89} & \gradientA{73} & \gradientA{57} \\
		\href{https://gymnasium.farama.org/environments/mujoco/ant}{Ant}                                           & \gradientA{100}                    & \gradientA{88}  & \gradientA{75} & \gradientA{52} & \gradientA{35} \\
		\href{https://gymnasium.farama.org/environments/classic_control/cart_pole/}{Cartpole}                      & \gradientA{100}                    & \gradientA{81}  & \gradientA{80} & \gradientA{57} & \gradientA{40} \\
		\href{https://gymnasium.farama.org/environments/mujoco/half_cheetah}{Half Cheetah}                         & \gradientA{100}                    & \gradientA{95}  & \gradientA{76} & \gradientA{61} & \gradientA{43} \\
		\href{https://gymnasium.farama.org/environments/mujoco/hopper}{Hopper}                                     & \gradientA{100}                    & \gradientA{98}  & \gradientA{92} & \gradientA{69} & \gradientA{50} \\
		\href{https://gymnasium.farama.org/environments/mujoco/humanoid}{Humanoid}                                 & \gradientA{100}                    & \gradientA{70}  & \gradientA{45} & \gradientA{26} & \gradientA{9}  \\
		\href{https://gymnasium.farama.org/environments/mujoco/humanoid_standup}{Humanoid Standup}                 & \gradientA{100}                    & \gradientA{68}  & \gradientA{40} & \gradientA{22} & \gradientA{10} \\
		\href{https://gymnasium.farama.org/environments/mujoco/inverted_double_pendulum}{Inverted Double Pendulum} & \gradientA{100}                    & \gradientA{82}  & \gradientA{77} & \gradientA{63} & \gradientA{43} \\
		\href{https://gymnasium.farama.org/environments/mujoco/inverted_pendulum}{Inverted Pendulum}               & \gradientA{100}                    & \gradientA{88}  & \gradientA{87} & \gradientA{69} & \gradientA{54} \\
		\href{https://gymnasium.farama.org/environments/classic_control/pendulum/}{Pendulum}                       & \gradientA{100}                    & \gradientA{100} & \gradientA{95} & \gradientA{82} & \gradientA{64} \\
		\href{https://gymnasium.farama.org/environments/mujoco/pusher}{Pusher}                                     & \gradientA{100}                    & \gradientA{95}  & \gradientA{85} & \gradientA{63} & \gradientA{41} \\
		\href{https://gymnasium.farama.org/environments/mujoco/reacher}{Reacher}                                   & \gradientA{100}                    & \gradientA{90}  & \gradientA{81} & \gradientA{64} & \gradientA{43} \\
		\href{https://gymnasium.farama.org/environments/mujoco/swimmer}{Swimmer}                                   & \gradientA{100}                    & \gradientA{91}  & \gradientA{79} & \gradientA{62} & \gradientA{39} \\
		\href{https://gymnasium.farama.org/environments/mujoco/walker2d}{Walker2d}                                 & \gradientA{100}                    & \gradientA{99}  & \gradientA{91} & \gradientA{72} & \gradientA{48} \\
		\bottomrule
	\end{tabular}
	\newline
	\caption{\label{tab:experiment-efficiency}Efficiency percentiles calculated as measured throughput divided by ideal throughout.}
\end{table}

We used \textit{throughput} (number of state-action interactions per second) as our primary metric. One would expect that as more CPU workers are added, the \textit{throughput} of state-action interactions would increase linearly.  In fact, that is exactly what we see, increasing the number of CPU workers on a Ray Cluster linearly scales the throughput.  Each experiment was run four times to get the mean and standard deviation.  For each CPU configuration, we calculated the throughput speedup factor.  We also provide an estimated efficiency percentage calculated as the measured throughput divided by the estimated ideal throughput. In instances where the ratio of measured throughout to ideal throughput is above 100\% due to experiment variance, we have rounded it down to 100\%.  Table~\ref{tab:experiment-speedup} and table~\ref{tab:experiment-efficiency} show experimental results.  Full details can be found in table~\ref{tab:experiment-config-1} and table~\ref{tab:experiment-config-2} in the appendix.


As more CPU workers were added the efficiency ratios decreased due to communication costs.  For some MuJoCo environments (i.e. Humanoid and Humanoid Stand-up), the communication costs outweighed the benefits of adding more CPU workers as they scaled.  But for the vast majority of the environments tested, increasing CPU workers increased throughput at a reasonable rate.


\section{Conclusion}\label{sec:conclusion}
Syndeo simultaneously addresses the issues of \textit{portability}, \textit{scalability}, and \textit{security} for HPC computing.  The design is portable because both the scheduler and the algorithms are containerized meaning it does not depend on the host system's hardware, scheduler, or installed libraries to operate.  The process is scalable because computing nodes can join the scheduler with just a copy of the container, a temporary writable directory, and network IP communication.  The process is secure because users are forced to operate with unprivileged profiles meaning administrators control the access permissions.


\section{Acknowledgements}
Special thanks to Albert Reuther, Chansup Byun, and Charles Yee from MIT Lincoln Laboratory Super Computing (LLSC)\cite{reuther2018interactive}.  Their insightful knowledge of how Slurm operates was essential in getting this project working.

\bibliographystyle{ieeetr}

\bibliography{main.bib}

\onecolumn 
\begin{appendices}
	\section{Experimental Results}
	\begin{figure*}[h]
		\centering
		\includegraphics[width=16cm]{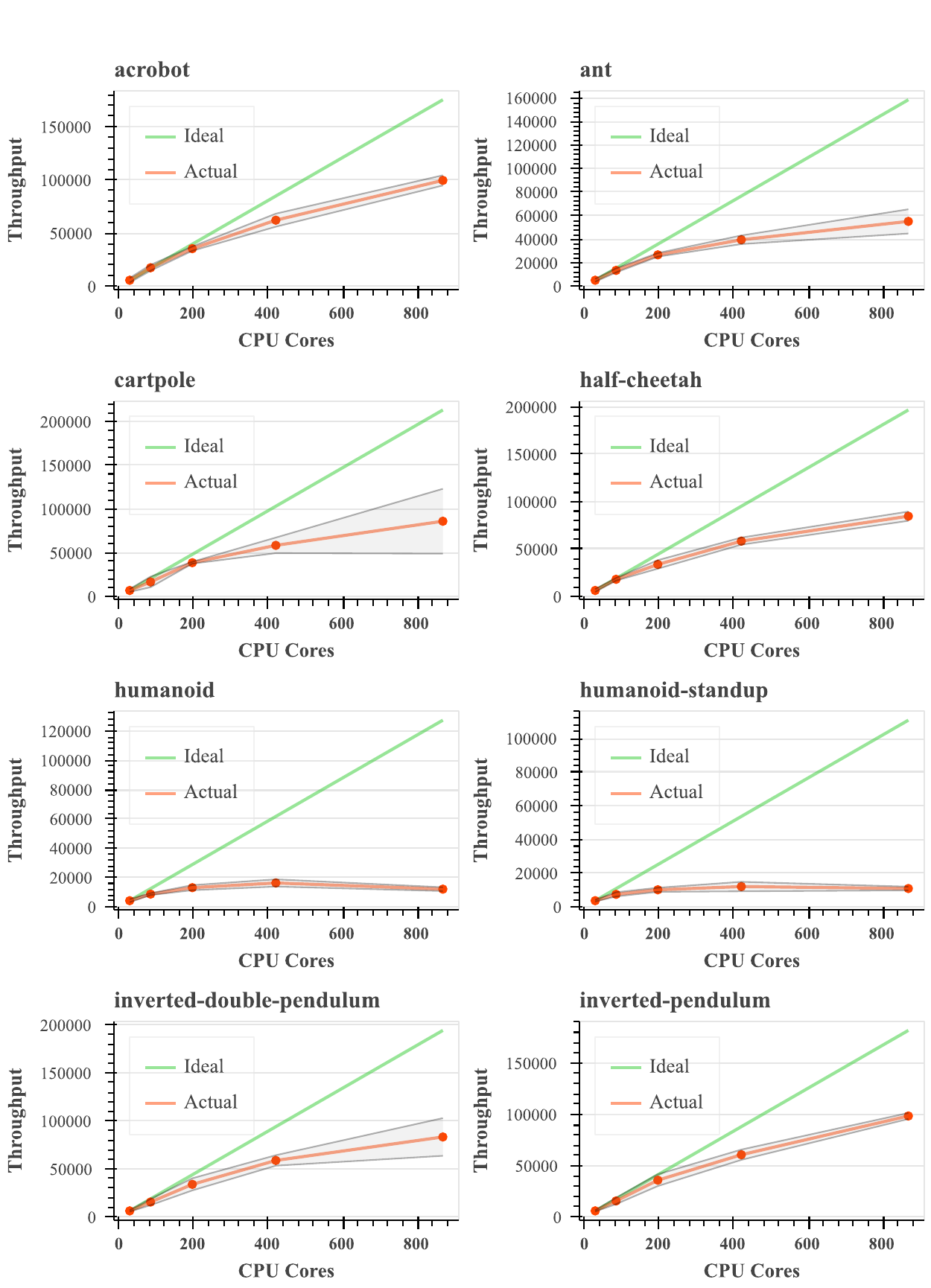}
		\caption{\label{fig:performance-line-0}Ideal versus measured performance with 2$\sigma$ standard deviation bands.  Here we assume that ideal is an extrapolation of the 28 CPU worker throughput.  As more CPU workers are added, the communication costs increase which degrades performance.}
	\end{figure*}

	\begin{figure*}[h]
		\centering
		\includegraphics[width=16cm]{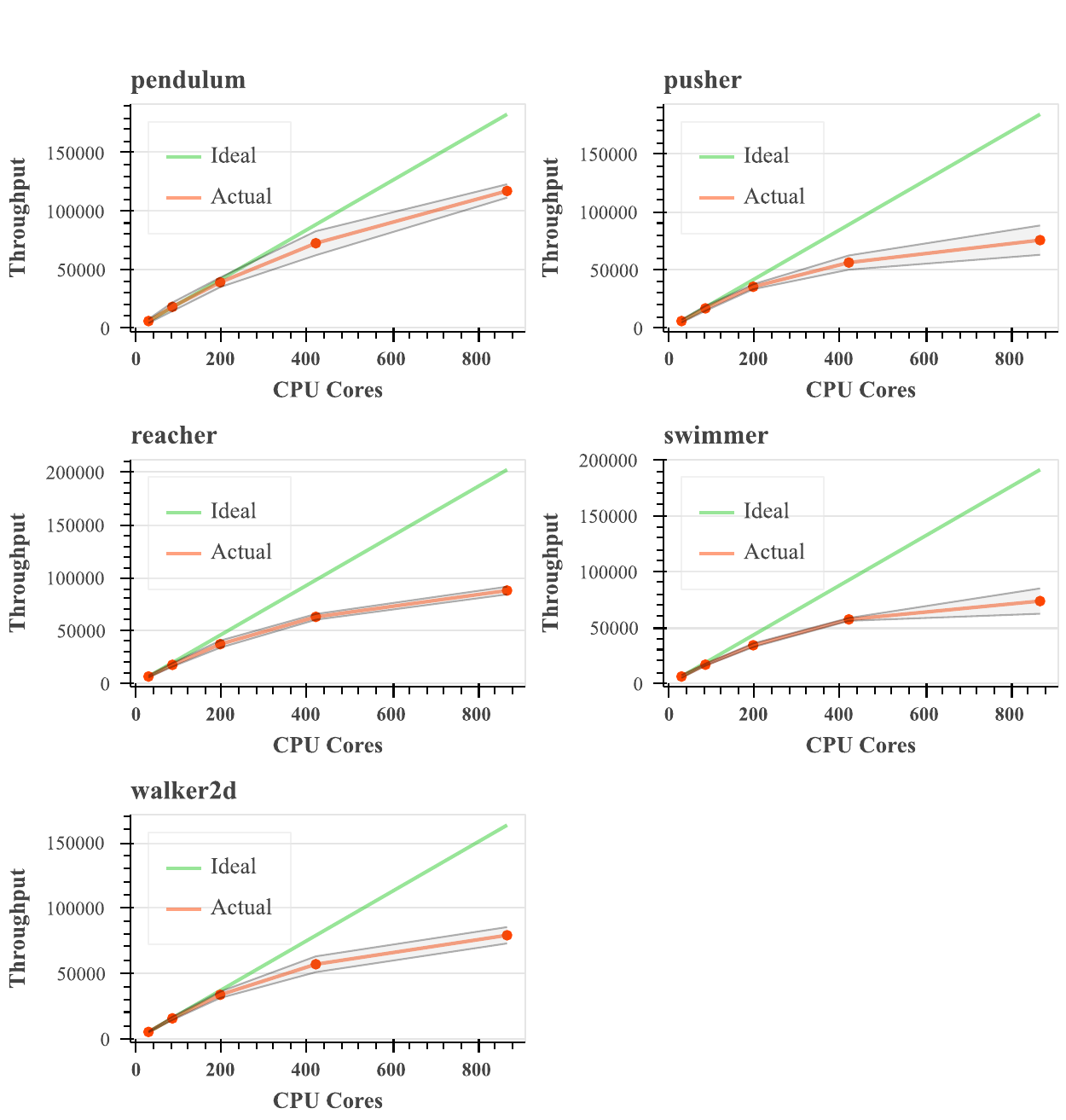}
		\caption{\label{fig:performance-line-4}Ideal versus measured performance with 2$\sigma$ standard deviation bands.  Here we assume that ideal is an extrapolation of the 28 CPU worker throughput.  As more CPU workers are added, the communication costs increase which degrades performance.}
	\end{figure*}

	\begin{table*}[h]
		\centering
		\fontfamily{ppl}\selectfont
		\resizebox{\textwidth}{!}{%
			\begin{tabular}{l|l|ll|ll|ll|lll}
				\toprule
				                                     &                                                     & \multicolumn{2}{c|}{\# Nodes} & \multicolumn{2}{c|}{Hardware} & \multicolumn{2}{c|}{Throughput} & \multicolumn{3}{c}{Speedup}                                                                       \\
				Environment                          & Processor Type                                      & Head                          & Workers                       & CPUs                            & GPUs                        & Mean        & Std Dev        & Ideal & Actual   & Actual/Ideal (\%) \\
				\toprule
				\multirow{5}{5em}{Acrobot}           & \multirow{5}{7em}{Xeon E5-2683 2.0Ghz - 256 GB RAM} & 1                             & 1                             & 28                              & 0                           & $\mu$=5656  & $\sigma$=1063  & 1x    & $\sim$1  & \gradientA{100}   \\
				                                     &                                                     & 1                             & 3                             & 84                              & 0                           & $\mu$=17391 & $\sigma$=1350  & 3x    & $\sim$3  & \gradientA{100}   \\
				                                     &                                                     & 1                             & 7                             & 196                             & 0                           & $\mu$=35370 & $\sigma$=1008  & 7x    & $\sim$6  & \gradientA{89}    \\
				                                     &                                                     & 1                             & 15                            & 420                             & 0                           & $\mu$=62134 & $\sigma$=3070  & 15x   & $\sim$11 & \gradientA{73}    \\
				                                     &                                                     & 1                             & 31                            & 868                             & 0                           & $\mu$=99540 & $\sigma$=2402  & 31x   & $\sim$18 & \gradientA{57}    \\
				\midrule
				\multirow{5}{5em}{Ant}               & \multirow{5}{7em}{Xeon E5-2683 2.0Ghz - 256 GB RAM} & 1                             & 1                             & 28                              & 0                           & $\mu$=5106  & $\sigma$=597   & 1x    & $\sim$1  & \gradientA{100}   \\
				                                     &                                                     & 1                             & 3                             & 84                              & 0                           & $\mu$=13492 & $\sigma$=866   & 3x    & $\sim$3  & \gradientA{88}    \\
				                                     &                                                     & 1                             & 7                             & 196                             & 0                           & $\mu$=26650 & $\sigma$=776   & 7x    & $\sim$5  & \gradientA{75}    \\
				                                     &                                                     & 1                             & 15                            & 420                             & 0                           & $\mu$=39468 & $\sigma$=1836  & 15x   & $\sim$8  & \gradientA{52}    \\
				                                     &                                                     & 1                             & 31                            & 868                             & 0                           & $\mu$=55055 & $\sigma$=5142  & 31x   & $\sim$11 & \gradientA{35}    \\
				\midrule
				\multirow{5}{5em}{Cartpole}          & \multirow{5}{7em}{Xeon E5-2683 2.0Ghz - 256 GB RAM} & 1                             & 1                             & 28                              & 0                           & $\mu$=6876  & $\sigma$=810   & 1x    & $\sim$1  & \gradientA{100}   \\
				                                     &                                                     & 1                             & 3                             & 84                              & 0                           & $\mu$=16634 & $\sigma$=3025  & 3x    & $\sim$2  & \gradientA{81}    \\
				                                     &                                                     & 1                             & 7                             & 196                             & 0                           & $\mu$=38618 & $\sigma$=531   & 7x    & $\sim$6  & \gradientA{80}    \\
				                                     &                                                     & 1                             & 15                            & 420                             & 0                           & $\mu$=58432 & $\sigma$=4457  & 15x   & $\sim$8  & \gradientA{57}    \\
				                                     &                                                     & 1                             & 31                            & 868                             & 0                           & $\mu$=85992 & $\sigma$=18533 & 31x   & $\sim$13 & \gradientA{40}    \\

				\midrule
				\multirow{5}{5em}{Half Cheetah}      & \multirow{5}{7em}{Xeon E5-2683 2.0Ghz - 256 GB RAM} & 1                             & 1                             & 28                              & 0                           & $\mu$=6343  & $\sigma$=752   & 1x    & $\sim$1  & \gradientA{100}   \\
				                                     &                                                     & 1                             & 3                             & 84                              & 0                           & $\mu$=18117 & $\sigma$=689   & 3x    & $\sim$3  & \gradientA{95}    \\
				                                     &                                                     & 1                             & 7                             & 196                             & 0                           & $\mu$=33788 & $\sigma$=2259  & 7x    & $\sim$5  & \gradientA{76}    \\
				                                     &                                                     & 1                             & 15                            & 420                             & 0                           & $\mu$=58362 & $\sigma$=1879  & 15x   & $\sim$9  & \gradientA{61}    \\
				                                     &                                                     & 1                             & 31                            & 868                             & 0                           & $\mu$=84636 & $\sigma$=2443  & 31x   & $\sim$13 & \gradientA{43}    \\

				\midrule
				\multirow{5}{5em}{Hopper}            & \multirow{5}{7em}{Xeon E5-2683 2.0Ghz - 256 GB RAM} & 1                             & 1                             & 28                              & 0                           & $\mu$=5505  & $\sigma$=466   & 1x    & $\sim$1  & \gradientA{100}   \\
				                                     &                                                     & 1                             & 3                             & 84                              & 0                           & $\mu$=16218 & $\sigma$=1211  & 3x    & $\sim$3  & \gradientA{98}    \\
				                                     &                                                     & 1                             & 7                             & 196                             & 0                           & $\mu$=35266 & $\sigma$=1449  & 7x    & $\sim$6  & \gradientA{92}    \\
				                                     &                                                     & 1                             & 15                            & 420                             & 0                           & $\mu$=56657 & $\sigma$=1396  & 15x   & $\sim$10 & \gradientA{69}    \\
				                                     &                                                     & 1                             & 31                            & 868                             & 0                           & $\mu$=85880 & $\sigma$=3545  & 31x   & $\sim$16 & \gradientA{50}    \\

				\midrule
				\multirow{5}{5em}{Humanoid}          & \multirow{5}{7em}{Xeon E5-2683 2.0Ghz - 256 GB RAM} & 1                             & 1                             & 28                              & 0                           & $\mu$=4108  & $\sigma$=514   & 1x    & $\sim$1  & \gradientA{100}   \\
				                                     &                                                     & 1                             & 3                             & 84                              & 0                           & $\mu$=8579  & $\sigma$=326   & 3x    & $\sim$2  & \gradientA{70}    \\
				                                     &                                                     & 1                             & 7                             & 196                             & 0                           & $\mu$=12991 & $\sigma$=865   & 7x    & $\sim$3  & \gradientA{45}    \\
				                                     &                                                     & 1                             & 15                            & 420                             & 0                           & $\mu$=16211 & $\sigma$=1239  & 15x   & $\sim$4  & \gradientA{26}    \\
				                                     &                                                     & 1                             & 31                            & 868                             & 0                           & $\mu$=11969 & $\sigma$=650   & 31x   & $\sim$3  & \gradientA{9}     \\

				\midrule
				\multirow{5}{5em}{Humanoid Stand-up} & \multirow{5}{7em}{Xeon E5-2683 2.0Ghz - 256 GB RAM} & 1                             & 1                             & 28                              & 0                           & $\mu$=3573  & $\sigma$=252   & 1x    & $\sim$1  & \gradientA{100}   \\
				                                     &                                                     & 1                             & 3                             & 84                              & 0                           & $\mu$=7321  & $\sigma$=627   & 3x    & $\sim$2  & \gradientA{68}    \\
				                                     &                                                     & 1                             & 7                             & 196                             & 0                           & $\mu$=10036 & $\sigma$=613   & 7x    & $\sim$3  & \gradientA{40}    \\
				                                     &                                                     & 1                             & 15                            & 420                             & 0                           & $\mu$=11950 & $\sigma$=1408  & 15x   & $\sim$3  & \gradientA{22}    \\
				                                     &                                                     & 1                             & 31                            & 868                             & 0                           & $\mu$=10794 & $\sigma$=553   & 31x   & $\sim$3  & \gradientA{10}    \\
			\end{tabular}}
		\newline
		\caption{\label{tab:experiment-config-1}Throughput is the number of state-action interactions collected per second.  The speed column shows the relative increase in throughput compared with the original 28 CPU workers.  Ray Clusters were set up with one scheduler node and one or more worker nodes.  Only the CPUs of the worker nodes are listed in the table above.}
	\end{table*}

	\begin{table*}[h]
		\centering
		\fontfamily{ppl}\selectfont
		\resizebox{\textwidth}{!}{%
			\begin{tabular}{l|l|ll|ll|ll|lll}
				\toprule
				                                            &                                                     & \multicolumn{2}{c|}{\# Nodes} & \multicolumn{2}{c|}{Hardware} & \multicolumn{2}{c|}{Throughput} & \multicolumn{3}{c}{Speedup}                                                                       \\
				Environment                                 & Processor Type                                      & Head                          & Workers                       & CPUs                            & GPUs                        & Mean         & Std Dev       & Ideal & Actual   & Actual/Ideal (\%) \\
				\toprule
				\multirow{5}{5em}{Inverted Double Pendulum} & \multirow{5}{7em}{Xeon E5-2683 2.0Ghz - 256 GB RAM} & 1                             & 1                             & 28                              & 0                           & $\mu$=6265   & $\sigma$=388  & 1x    & $\sim$1  & \gradientA{100}   \\
				                                            &                                                     & 1                             & 3                             & 84                              & 0                           & $\mu$=15470  & $\sigma$=1612 & 3x    & $\sim$2  & \gradientA{82}    \\
				                                            &                                                     & 1                             & 7                             & 196                             & 0                           & $\mu$=33887  & $\sigma$=3162 & 7x    & $\sim$5  & \gradientA{77}    \\
				                                            &                                                     & 1                             & 15                            & 420                             & 0                           & $\mu$=58794  & $\sigma$=2731 & 15x   & $\sim$9  & \gradientA{63}    \\
				                                            &                                                     & 1                             & 31                            & 868                             & 0                           & $\mu$=83240  & $\sigma$=9815 & 31x   & $\sim$13 & \gradientA{43}    \\

				\midrule
				\multirow{5}{5em}{Inverted Pendulum}        & \multirow{5}{7em}{Xeon E5-2683 2.0Ghz - 256 GB RAM} & 1                             & 1                             & 28                              & 0                           & $\mu$=5864   & $\sigma$=300  & 1x    & $\sim$1  & \gradientA{100}   \\
				                                            &                                                     & 1                             & 3                             & 84                              & 0                           & $\mu$=15555  & $\sigma$=1550 & 3x    & $\sim$3  & \gradientA{88}    \\
				                                            &                                                     & 1                             & 7                             & 196                             & 0                           & $\mu$=35779  & $\sigma$=2868 & 7x    & $\sim$6  & \gradientA{87}    \\
				                                            &                                                     & 1                             & 15                            & 420                             & 0                           & $\mu$=60589  & $\sigma$=2515 & 15x   & $\sim$10 & \gradientA{69}    \\
				                                            &                                                     & 1                             & 31                            & 868                             & 0                           & $\mu$=98372  & $\sigma$=1471 & 31x   & $\sim$17 & \gradientA{54}    \\

				\midrule
				\multirow{5}{5em}{Pendulum}                 & \multirow{5}{7em}{Xeon E5-2683 2.0Ghz - 256 GB RAM} & 1                             & 1                             & 28                              & 0                           & $\mu$=5895   & $\sigma$=845  & 1x    & $\sim$1  & \gradientA{100}   \\
				                                            &                                                     & 1                             & 3                             & 84                              & 0                           & $\mu$=18100  & $\sigma$=1844 & 3x    & $\sim$3  & \gradientA{100}   \\
				                                            &                                                     & 1                             & 7                             & 196                             & 0                           & $\mu$=39065  & $\sigma$=2060 & 7x    & $\sim$7  & \gradientA{95}    \\
				                                            &                                                     & 1                             & 15                            & 420                             & 0                           & $\mu$=72493  & $\sigma$=5152 & 15x   & $\sim$12 & \gradientA{82}    \\
				                                            &                                                     & 1                             & 31                            & 868                             & 0                           & $\mu$=117197 & $\sigma$=2874 & 31x   & $\sim$20 & \gradientA{64}    \\

				\midrule
				\multirow{5}{5em}{Pusher}                   & \multirow{5}{7em}{Xeon E5-2683 2.0Ghz - 256 GB RAM} & 1                             & 1                             & 28                              & 0                           & $\mu$=5939   & $\sigma$=592  & 1x    & $\sim$1  & \gradientA{100}   \\
				                                            &                                                     & 1                             & 3                             & 84                              & 0                           & $\mu$=16879  & $\sigma$=1046 & 3x    & $\sim$3  & \gradientA{95}    \\
				                                            &                                                     & 1                             & 7                             & 196                             & 0                           & $\mu$=35535  & $\sigma$=1066 & 7x    & $\sim$6  & \gradientA{85}    \\
				                                            &                                                     & 1                             & 15                            & 420                             & 0                           & $\mu$=56371  & $\sigma$=3067 & 15x   & $\sim$9  & \gradientA{63}    \\
				                                            &                                                     & 1                             & 31                            & 868                             & 0                           & $\mu$=75679  & $\sigma$=6325 & 31x   & $\sim$13 & \gradientA{41}    \\

				\midrule
				\multirow{5}{5em}{Reacher}                  & \multirow{5}{7em}{Xeon E5-2683 2.0Ghz - 256 GB RAM} & 1                             & 1                             & 28                              & 0                           & $\mu$=6521   & $\sigma$=428  & 1x    & $\sim$1  & \gradientA{100}   \\
				                                            &                                                     & 1                             & 3                             & 84                              & 0                           & $\mu$=17555  & $\sigma$=833  & 3x    & $\sim$3  & \gradientA{90}    \\
				                                            &                                                     & 1                             & 7                             & 196                             & 0                           & $\mu$=37128  & $\sigma$=1685 & 7x    & $\sim$6  & \gradientA{81}    \\
				                                            &                                                     & 1                             & 15                            & 420                             & 0                           & $\mu$=62964  & $\sigma$=1364 & 15x   & $\sim$10 & \gradientA{64}    \\
				                                            &                                                     & 1                             & 31                            & 868                             & 0                           & $\mu$=87839  & $\sigma$=1839 & 31x   & $\sim$13 & \gradientA{43}    \\

				\midrule
				\multirow{5}{5em}{Swimmer}                  & \multirow{5}{7em}{Xeon E5-2683 2.0Ghz - 256 GB RAM} & 1                             & 1                             & 28                              & 0                           & $\mu$=6168   & $\sigma$=605  & 1x    & $\sim$1  & \gradientA{100}   \\
				                                            &                                                     & 1                             & 3                             & 84                              & 0                           & $\mu$=16767  & $\sigma$=447  & 3x    & $\sim$3  & \gradientA{91}    \\
				                                            &                                                     & 1                             & 7                             & 196                             & 0                           & $\mu$=34129  & $\sigma$=820  & 7x    & $\sim$6  & \gradientA{79}    \\
				                                            &                                                     & 1                             & 15                            & 420                             & 0                           & $\mu$=57256  & $\sigma$=635  & 15x   & $\sim$9  & \gradientA{62}    \\
				                                            &                                                     & 1                             & 31                            & 868                             & 0                           & $\mu$=73660  & $\sigma$=5674 & 31x   & $\sim$12 & \gradientA{39}    \\

				\midrule
				\multirow{5}{5em}{Walker2d}                 & \multirow{5}{7em}{Xeon E5-2683 2.0Ghz - 256 GB RAM} & 1                             & 1                             & 28                              & 0                           & $\mu$=5264   & $\sigma$=208  & 1x    & $\sim$1  & \gradientA{100}   \\
				                                            &                                                     & 1                             & 3                             & 84                              & 0                           & $\mu$=15618  & $\sigma$=597  & 3x    & $\sim$3  & \gradientA{99}    \\
				                                            &                                                     & 1                             & 7                             & 196                             & 0                           & $\mu$=33623  & $\sigma$=1224 & 7x    & $\sim$6  & \gradientA{91}    \\
				                                            &                                                     & 1                             & 15                            & 420                             & 0                           & $\mu$=56953  & $\sigma$=3029 & 15x   & $\sim$11 & \gradientA{72}    \\
				                                            &                                                     & 1                             & 31                            & 868                             & 0                           & $\mu$=79085  & $\sigma$=3128 & 31x   & $\sim$15 & \gradientA{48}    \\
			\end{tabular}}
		\newline
		\caption{\label{tab:experiment-config-2}Throughput is the number of state-action interactions collected per second.  The speed column shows the relative increase in throughput compared with the original 28 CPU workers.  Ray Clusters were set up with one scheduler node and one or more worker nodes.  Only the CPUs of the worker nodes are listed in the table above.}
	\end{table*}
\end{appendices}

\end{document}